\documentclass[11pt]{article} 

\usepackage{amssymb,amsmath,epsfig,color}% Include figure files
\usepackage{framed}
\newtheorem{theorem}{Theorem}
\newtheorem{remark}[theorem]{Remark}

\newcommand{\rem}[1]{}
\def\contract{\makebox[1.2em][c]{\mbox{\rule{.6em}
{.01truein}\rule{.01truein}{.6em}}}}
\newcommand{\bfi}{\bfseries\itshape}
\newcommand{\pa}{\partial}
\newcommand{\itb}{\color{blue}\itshape}
\newcommand{\bg}{\boldsymbol{g}}
\newcommand{\comment}[1]{\vspace{1 mm}\par 
\marginpar{\large\underline{}}\noindent
\framebox{\begin{minipage}[c]{1 \textwidth}
{\itb #1} \end{minipage}}\vspace{1 mm}\par}

\begin{document}
\title{Geometry of Vlasov kinetic moments:\\ a bosonic Fock space for the symmetric Schouten bracket}
\author{\vspace{2mm}
John Gibbons$^{1}$, Darryl D. Holm$^{1,\,2}$ and Cesare Tronci$^{1,3}\!$\\
\vspace{-.2cm}
\\
{\footnotesize $^1$ \it Department of Mathematics, Imperial College London, London SW7 2AZ, UK}\\
{\footnotesize $^2$ \it Computer and Computational Science Division, Los Alamos National Laboratory, Los Alamos, NM, 87545 USA} \\
{\footnotesize $^3$\,\it TERA Foundation for Oncological Hadrontherapy,
11 V. Puccini, Novara 28100, Italy}
\\ \\
}
\date{}

\maketitle

\vspace{-1cm}
\begin{abstract}\noindent
The dynamics of Vlasov kinetic moments is shown to be Lie-Poisson on the dual Lie algebra of symmetric contravariant tensor fields. The corresponding Lie bracket is identified with the symmetric Schouten bracket and the moment
Lie algebra is related with a bundle of bosonic Fock spaces, where creation
and annihilation operators are used to construct the cold plasma closure. Kinetic moments are also shown to define a momentum map, which is infinitesimally equivariant. This momentum map is the dual of a Lie algebra homomorphism, defined through the Schouten bracket. Finally the moment Lie-Poisson bracket is extended to anisotropic interactions.
\end{abstract}

\tableofcontents

%%%%%%%%%%%%%%%%

\section{Geometric methods in kinetic theory}
\label{kinetic theory-sec}

\paragraph{Liouville equation.}
The equations of kinetic theory in non-equilibrium statistical mechanics
always describe the evolution of a probability distribution in phase space.
The most fundamental kinetic equation is the {\bfi Liouville equation} describing
the evolution of an $N$-particle distribution $\rho(z_1,\dots,z_N)$, where
$z_i=(q_i,p_i)$ is the phase space coordinate of the $i$-th particle. Given
the conservation of particles and the Hamiltonian nature
of their dynamics, the preservation of the symplectic
(phase space) volume assures that the equation for $\rho$ is written as the
advection equation $\pa_{t\,} \rho\,+\,\left\{\rho,H\right\}=0$,
where the Hamiltonian $H$ depends on all particles and $\{\cdot,\cdot\}$
is the canonical Poisson bracket. 

In geometric terms the Liouville equation is a Lie-Poisson equation \cite{MaRa99}
\rem{ %%%%%%%%%%%%%%%%% BEGIN REM %%%%%%%%%%%%%%%%%%%%%%%%%%%%%%%
,
i.e. a Hamiltonian system defined on a dual Lie algebra $\mathfrak{g}^*$
arising from the following poisson bracket
\[
\{F,G\}=\left\langle\mu,\left[\frac{\delta F}{\delta \mu},\frac{\delta G}{\delta \mu}\right]\right\rangle
\]
where $\mu\in\mathfrak{g}^*$, and $\delta F/\delta \mu$ denotes functional
(Fr\'echet) derivative. The corresponding equation is written as $\dot{\mu}+{\rm
ad}^*_{\,\delta \mathcal{H}/\delta \mu}\,\mu=0$.
} %%%%%%%%%%%%%%%%% END REM %%%%%%%%%%%%%%%%%%%%%%%%%%%%%%%
 on the
group of canonical transformations (symplectomorphisms) on the $N$-particle
 phase space $T^*Q\times\dots\times T^*Q$ ($N$ times) \cite{MaMoWe1984}. The Lie algebra of the symplectic group consists
of Hamiltonian vector fields ${\bf X}_H\in\mathfrak{X}_\text{can}$ and is identified with the Poisson
algebra of generating functions $H\in\mathcal{F}$, via the usual Lie algebra anti-homomorphism
${\bf X}_H[f]=\{f,H\}$. Thus the dual Lie algebra is identified with the
space of densities (distributions) $\rho\in\mathcal{F}^*\simeq\text{Den}$, which undergo Lie-Poisson dynamics. 

\paragraph{The BBGKY hierarchy.}
In the search for approximate descriptions of the Liouville dynamics, it is common to introduce global quantities that integrate out the information on some of the particles (usually most of them). In particular one defines a $n$-particle distribution as
\[
f_n(z_1,\dots,z_n,t):=\int \rho(z_1,\dots,z_N)\,{\rm d}z_{n+1}\dots{\rm d}z_N
\,.
\]
These quantities are called {\bfi BBGKY moments} and their equations
constitute a hierarchy of equations known as {BBGKY hierarchy}, or {\it Bogoliubov-Born-Green-Kirkwood-Yvon equations}.
This hierarchy is rather complicated, although Marsden, Morrison and Weinstein
\cite{MaMoWe1984} have shown that this is also a Lie-Poisson system, which
is defined
on the Lie algebra of canonical transformations that are symmetric with respect
to their arguments. Thus the operation of taking the BBGKY moments is a Poisson map. More particularly, the reason why this happens is that BBGKY moments are a particular kind of {\bfi momentum map} \cite{MaRa99}.

\paragraph{The Vlasov equation.}
Suitable approximations on the equation for the single particle distribution $f:=f_1$ (the first-order BBGKY moment) yield the {\it Boltzmann equation}
\cite{Bo95},
which we write simply as
\[
\frac{\pa f}{\pa t}+\left\{f,H\right\}=\left(\frac{\pa f}{\pa t}\right)_{coll}
\,,
\]
where $H$ is now the single-particle Hamiltonian. The right hand
side collects the information on pairwise collisions among particles. An important approximation of the Boltzmann equation is the {\it Fokker-Planck equation} \cite{Fokker-Plank1931}, arising from the stochastic hypothesis of brownian motion.
\rem{ %%%%%%%%%%%%%%%%% BEGIN REM %%%%%%%%%%%%%%%%%%%%%%%%%
 that leads to the substitution of the collision integral with the drift-diffusion term
\[
\left(\frac{\pa f}{\pa t}\right)_{coll\!}\!=\,\frac{\pa}{\pa p}\left(\gamma\, p f+D\,\frac{\pa
f}{\pa p}\right)
\,.
\]
} %%%%%%%%%%%%%%%%% ENd REM %%%%%%%%%%%%%%%%%%%%%%%%%
However, in many contexts it is possible to neglect the effects of collisions,
so that
\begin{equation}\label{Vlasov-eq}
\frac{\pa f}{\pa t}+\left\{f,H\right\}=0
\,.
\end{equation}
This is called {\bfi Vlasov equation} \cite{Vl1961} and its underlying geometric
structure has been widely investigated \cite{Lo58,WeMo,MaWe,MaWeRaScSp,CeHoHoMa1998}.
In particular, Marsden, Weinstein and collaborators \cite{MaWe,MaWeRaScSp} have shown that this equation possesses a Lie-Poisson structure on the group of canonical transformations of the single-particle phase space $T^*Q$. The explicit expression of the Lie-Poisson bracket is 
\begin{equation}
\left\{F,G\right\}_{V}[f]\,=\int\hspace{-3mm}\int \!f(q,p,t)\left\{\frac{\delta F}{\delta f},\frac{\delta
G}{\delta f}\right\}{\rm d}q\,{\rm d}p
\,,
\label{VLP}
\end{equation}
where the index $V$ will be used below to avoid confusion in the notation.
\rem{ %%%%%%%%%%%%%%%%%%%%%%% BEGIN REM %%%%%%%%%%%%%%%%%%%%%%%%%%%%%%
Even when this equation is coupled with the Maxwell equations ({\it Maxwell-Vlasov system}), the geometric structure persists \cite{MaWe,MaWeRaScSp}. Analogous
arguments also hold in the Lagrangian formulation \cite{CeHoHoMa1998}, which
has been first proposed in \cite{Lo58}.
} %%%%%%%%%%%%%%%%%%%%%%% END REM %%%%%%%%%%%%%%%%%%%%%%%%%%%%%%

\smallskip
In the general case, the Vlasov density $f$ depends on seven variables: six phase space coordinates plus time. This indicates that the Vlasov equation
is still rather complicated and it may be convenient to find suitable approximations. To this purpose, one introduces the {\bfi moments of the Vlasov distribution}. The use of moments in kinetic theory was introduced by Chapmann and Enskog \cite{Chapman1960}, who formulated their closure of the Boltzmann equation yielding the equations of fluid mechanics. This result showed the power of
using moments and the mathematical properties of moments have been widely investigated since that time.

\paragraph{Vlasov moments I: statistical moments} 
The investigation of the geometric properties of Vlasov moments mainly arose in two very different
contexts, particle beam dynamics and shallow water equations. However it
is important to distinguish between two different classes of moments: {\it
statistical moments} and {\it kinetic moments}. {\bfi Statistical moments}
are defined as
\[
\mathcal{M}_{n,\widehat{n}}(t):=\int\hspace{-3mm}\int p^{n\,} q^{\widehat{n}}\,
f(q,p,t)\,{\rm d}q\,{\rm d}p
\,.
\]
These quantities first arose in the study of particle beam dynamics \cite{Ch83,Ch90,LyOv88}
from the observation that the beam {\it emittance} $\epsilon:=\left(\mathcal{M}_{0,2}\,\mathcal{M}_{2,0}-\mathcal{M}_{1,1}^{\,2}\right)^{1/2}$
is a laboratory parameter, which is also invariant. In particular, Channell, Holm, Lysenko and Scovel \cite{Ch90,Ch95,HoLySc1990,LyPa97}
were the first to consider the Lie-Poisson structure of the moments, whose
explicit expression is
\[
\{\,F\,,\,G\,\}
\quad\!=
\sum_{\widehat{m},m=0\,}^\infty 
\sum_{\,\widehat{n},n=0}^\infty
\left[
\frac{\pa F}{\pa \mathcal{M}_{\widehat{m},m}}\,
\big(\widehat{m}\,m-\widehat{n}\,n\big)\,
\frac{\pa G}{\pa \mathcal{M}_{\widehat{n},n}}
\right]
\,
\mathcal{M}_{\,\widehat{m}+\widehat{n}-1,\,m+n-1}
\,.
\]
This
geometric framework allowed the systematic construction of symplectic moment invariants
in \cite{HoLySc1990}, a question that was also pursued by Dragt and collaborators
in \cite{DrNeRa92}. Special truncations have been studied also by Scovel and Weinstein in \cite{ScWe}.

\paragraph{Vlasov moments II: kinetic moments.}
Another class of moments was used by Chapman and Enskog \cite{Chapman1960}, i.e. the {\bfi kinetic moments}
\begin{equation}\label{mom-def}
A_n(q,t):=\int p^n f(q,p,t)\,{\rm d}p
\,.
\end{equation}
The following discussion will refer to these quantities as simply ``moments'',
unless otherwise specified. 
\rem{ %%%%%%%%%%%%%%%%%%% BEGIN REM %%%%%%%%%%%%%%%%%%%%%%%%%%%%
The geometric properties of these moments first arose when Gibbons \cite{Gi1981} recognized that these Vlasov moments are equivalent to the variables previously introduced by Benney \cite{Be1973}, in the context of shallow water
waves. 
} %%%%%%%%%%%%%%%%%%% END REM %%%%%%%%%%%%%%%%%%%%%%%%%%%%
The Hamiltonian structure of these variables was found by Kupershmidt and Manin \cite{KuMa} in the context of Benney long
waves \cite{Be1973}. Later, Gibbons recognized that this is a Lie-Poisson structure  inherited from the Vlasov bracket \cite{Gi1981}. The relation between moments and the symplectic Lie algebra was also known to Lebedev \cite{Le1979}, although he did not recognize the connection with Vlasov dynamics. The Lie-Poisson structure for the moments is also called {\it Kupershmidt-Manin structure}
and is explicitly written as \cite{KuMa}
\begin{equation}\label{KMLP}
\{F,G\}=-\int\! A_{m+n-1}\left[n\,\frac{\delta F}{\delta A_n}\frac{\pa}{\pa q}\frac{\delta G}{\delta A_m}
-
m\,\frac{\delta G}{\delta A_m}\frac{\pa}{\pa q}\frac{\delta F}{\delta A_n}
\right]{\rm d}q
\,.
\end{equation}
The main theorem regarding moments is thus the following
\begin{theorem}[Gibbons \cite{Gi1981}] The process of taking moments of the
Vlasov distribution is a {\bfi Poisson map}, which takes the Vlasov
Lie-Poisson structure to the Kupershmidt-Manin bracket.
\end{theorem}
This argument was also used later by Gibbons, Holm and Kupershmidt \cite{GiHoKu1982,GiHoKu1983}
in obtaining the moment equations of chromohydrodynamics, the continuum theory of a Yang-Mills quark-gluon plasma.

The Lie-Poisson structure of moment dynamics remains an interesting research topic. Indeed, there are many
open questions, some of which are answered in the present paper. For example,
Poisson maps relating different Lie-Poisson structures are often understood
in terms of momentum maps: What is the corresponding momentum map for kinetic moments? And also, the moment Lie-Poisson
bracket defines a Lie algebra: What is its geometric interpretation? The last question is especially important in higher dimensions. Kupershmidt proposed \cite{Ku1987,Ku2005} a multi-index treatment, so that moments may be defined as
\begin{equation}
A_{\sigma}
\left(  
\mathbf{q},t\right)  
:=
{\int}
p^{\sigma}
f\left(\mathbf{q},\mathbf{p},t\right)
{\rm d}^3{\bf p}
\,,
\label{mom-def3D}
\end{equation}
where $\sigma$ is a multi-index such that $p^{\sigma} :=
p_{1}^{\sigma_{1}}\,p_{2}^{\sigma_{2}}\,p_{3}^{\sigma_{3}}$. The corresponding
Lie-Poisson structure is found to be
\begin{equation}\label{KM-mi}
\{F,G\}=-\,
\sum_{\sigma ,\rho }\sum_{j}
\int\!
A_{\sigma +\rho -1_{\hbox{\footnotesize\it j}}}\,
\left[
\sigma _{j}\,
\frac{\delta F}{\delta A_\sigma}\,
\frac{\partial}{\partial q^{j}}\frac{\delta G}{\delta A_\rho}
-
\rho _{j}\,\frac{\delta G}{\delta A_\rho}\,
\frac{\partial}{\partial q^{j}}
\frac{\delta F}{\delta A_\sigma}
\right]
{\rm d}^3{\bf q}
\,,
\end{equation}
where $1_{\hbox{\footnotesize\it j}}$ is a multi-index such that $(1_{\hbox{\footnotesize\it j}})_k=\delta_{jk}$ (all zero entries, except the $j$-th, which is equal to unity). However, this notation does not address the geometric interpretation of the moments. Are they differential forms, special tensors, or something else? \medskip

\noindent{\bf Plan.} Section \ref{Lie-Poisson-sec} of this paper shows that moments are symmetric tensor densities whose Lie bracket is defined by the Schouten concomitant \cite{Schouten}. Section \ref{mom map-sec} shows that moments define a momentum map and investigates the properties of this map. The moment Lie algebra is then shown in Section \ref{Fock-sec} to possess a Fock-space structure as for Bose-Einstein statistics, and its creation and annihilation operators are used to construct the cold-plasma closure. In Section \ref{anisotropy-sec} the moment treatment is extended to include anisotropic interactions in accord with \cite{GiHoKu1982,GiHoKu1983}.  Conclusions and open questions are summarized briefly in Section \ref{conclusion-sec}. 

\section{Lie-Poisson structure of kinetic moments}
\label{Lie-Poisson-sec}

The Vlasov kinetic moments have been shown \cite{Gi1981} to possess the Kupershmidt-Manin
structure \cite{KuMa} given in (\ref{KMLP}). This property holds in one dimension and the analogous result (\ref{KM-mi}) holds in higher dimensions, when the moments are defined as in (\ref{mom-def3D}). The procedure to obtain
the Lie Poisson structure for the moments begins with the Vlasov bracket (\ref{VLP}) and makes use of the chain-rule identity
\begin{equation}
\label{chain}
\frac{\delta F}{\delta f}=
\sum_{n=0}^\infty\,\frac{\delta A_n}{\delta f}\frac{\delta F}{\delta A_n}=
\sum_{n=0}^\infty\, p^{n\,}\frac{\delta F}{\delta A_n}
\,,
\end{equation}
for any functional $F[f]$. Direct substitution of this expression into equation
(\ref{VLP}) yields the Lie-Poisson structure for the moments (\ref{KMLP}).
In higher dimensions, the same argument holds upon replacing $A_n$ with $A_\sigma$. 

\subsection{Tensorial interpretation of the moments}
In one dimension, the moment Lie-Poisson structure identifies a Lie algebra
defined by the square bracket in (\ref{KMLP}), while the multi-index
notation yields the square bracket in (\ref{KM-mi}) in higher dimensions. The questions addressed in this section are: What Lie algebra is this? What is its geometric nature?
For these considerations, the moments are regarded as fiber integrals \cite{AbMaRa,QiTa2004} on $T^*_q Q$, the Vlasov distribution is taken to be a phase space density $f\,{\rm d}q\wedge{\rm d}p$ and the natural geometry of the cotangent bundle $T^*Q$ is used to identify the momentum coordinate $p$ with the canonical one form
$\theta=p\,{\rm d}q$. In this way the moments are defined as
\begin{equation}
A_n=\int_{T^*_q Q\!}\otimes^n\!\left(p\,{\rm d}q\right)f(q,p,t)\,{\rm d}q\wedge{\rm d}p=A_n(q,t)\otimes^{n+1\!}{\rm d}q
\,.
\end{equation}
Since moments belong to a dual Lie algebra, the dual variables of the
moments are tensor fields of the form $\beta_n=\beta_n(q,t)\otimes^{n\!}\pa_q$.
Therefore, moments  in one dimension are covariant tensor-densities dual to the space of contravariant tensor fields, endowed with the Lie bracket in equation (\ref{KMLP}). This
conclusion was obtained earlier in \cite{GiHoTr2007}. However,
in order to enrich this analysis we need to extend our arguments to higher dimensions. Following the same line of reasoning allows one to write moments in higher dimensions as
\begin{align}
A_\sigma\,=\int_{T^*_q Q\!}
\left(p_1\,{\rm d}q^1\right)^{\sigma_1}\!
\left(p_2\,{\rm d}q^2\right)^{\sigma_2}\!
\left(p_3\,{\rm d}q^3\right)^{\sigma_3}
f({\bf q,p},t)\,\,{\rm d}^3{\bf q}\wedge{\rm d}^3{\bf p}
%\\&
\,\,=\,
A_\sigma({\bf q},t)\left({\rm d}{\bf q}\right)^\sigma\otimes{\rm d}^3{\bf q}
\,,
\end{align}
where ${\rm d}^3{\bf q}$ is the volume element on $\mathbb{R}^3$.

At this point we prefer to forgo the multi-index notation and to adopt
the definition of the moments that is used in the physics literature \cite{GoZhSa80}. Hence, moments become defined as tensors of the form
\begin{equation}
A_n({\bf q},t)=\int{\bf p}^n\,f({\bf q,p},t)\,\,{\rm d}^3{\bf p}
\,,
\end{equation}
where ${\bf p}^n$ denotes the $n$-th tensor power. Inserting the basis and
extending to any number $N$ of spatial dimensions yields
\begin{align}\nonumber
A_n({\bf q}, t)&=\sum_{i=1}^\infty\int_{T_{\bf q}^*Q} \! \left(p_i\,{\rm d}q^i\right)^n f({\bf q,
p}, t)\,{\rm d}^N{\bf q}\wedge{\rm d}^N{\bf p}
\\
&=\sum_{i_1,\dots,i_n}\int_{T_{\bf q}^*Q}  p_{\,i_1}\dots p_{\,i_n} \,{\rm d}q^{i_1}\dots{\rm d}q^{i_n}\,f({\bf q,
p}, t)\,{\rm d}^N{\bf q}\wedge{\rm d}^N{\bf p}
\nonumber
\\
&=\sum_{i_1,\dots,i_n}\!
\big(A_n({\bf q}, t)\big)_{i_1\dots i_n}
\,{\rm d}q^{i_1}\otimes\dots\otimes{\rm d}q^{i_n}\otimes{\rm d}^N{\bf q}
\,.
\label{mom3D}
\end{align}
This definition identifies the moments with {\it symmetric} covariant
tensor-densities. Also, since contractions of the form ${\bf p}^n\!\contract\beta_n({\bf q})$ as in (\ref{chain}) must be symmetric under permutations of indices, the Lie algebra variable may be identified with {\it symmetric contravariant tensor fields}. 

%\rem{ %%%%%%%%%%%%%%%%%%%%%%%%%%%5 BEGIN REM %%%%%%%%%%%%%%%%%%%%%%%%%%%%%%%5
%\begin{framed}
The duality between covariant and contravariant  tensors can be
seen by the following argument. Take initially $\beta_n=\hat{S}\beta_n+\hat{A}\beta_n$ as a generic contravariant $n$-tensor, where $\hat{S}$ and $\hat{A}$ are the symmetrizing and anti-symmetrizing operators respectively: the contraction with the tensor power ${\bf p}^n=\hat{S}{\bf p}^n$ yields
\[
{\bf p}^n\!\contract \beta_n({\bf q})
=\hat{S}{\bf p}^n\!\contract \beta_n({\bf q})
={\bf p}^n\!\contract\,\hat{S}\, \beta_n({\bf q})
\,,
\]
since $\hat{S}=\hat{S}^*$ \cite{Shaw}. Thus, the antisymmetric part of $\beta_n$ does not contribute to the contraction and $\beta_n$ can be identified with its symmetric part
$\hat{S}\beta_n$. The same argument
may be used to identify the dual of symmetric covariant tensor densities with the
space of symmetric contravariant tensors. Indeed, one has 
$\langle A_n, \beta_n\rangle
=\langle \hat{S}A_n, \beta_n\rangle
=\langle A_n, \hat{S}\beta_n\rangle
$, 
so that again $\beta_n$ is identified with its symmetric part.
%\end{framed}
%} %%%%%%%%%%%%%%%%%%%%%%%%%%%5 END REM %%%%%%%%%%%%%%%%%%%%%%%%%%%%%%%5

\subsection{The Schouten concomitant for Vlasov kinetic moments}
At this point, since the tensor notation is used in the same fashion in any
number of dimensions, we continue to work in terms of the symmetric tensors
$A_n$. However the Lie-Poisson structure for the moments is well-known only in multi-index notation \cite{Ku1987,Ku2005}. What is the Lie-Poisson structure for the tensor moments? More particularly, the Lie algebra is now the space of symmetric
contravariant tensor fields. What is its Lie bracket? To answer this question, it suffices to insert
the expression
\begin{equation}
\frac{\delta F}{\delta f}=
\sum_{n=0}^\infty\,\sum_{\,i_1\dots i_n=1}^N\!\frac{\delta (A_n)_{i_1\dots i_n}}{\delta f}\,\frac{\delta F}{\delta (A_n)_{i_1\dots i_n}}
=
\sum_{n=0}^\infty\,\frac{\delta A_n}{\delta f}\contract\frac{\delta F}{\delta A_n}=
\sum_{n=0}^\infty\, {\bf p}^{n\!}\contract\frac{\delta F}{\delta A_n}
\end{equation}
into the Vlasov bracket (\ref{VLP}) to obtain the following Lie-Poisson structure
for the moments $A_n$
\begin{equation}
\left\{F,G\right\}=\,-
\sum_{n,m=0\,}^\infty
%\left\langle
\int\!
A_{m+n-1}\left[ n
\left(\frac{\delta F}{\delta A_n}\contract\,\nabla\right)\frac{\delta G}{\delta A_m}
-
m
\left(\frac{\delta G}{\delta A_m}\contract\,\nabla\right)\frac{\delta F}{\delta A_n}\right]
%\right\rangle
\,{\rm d}^N{\bf q}
\,,
\label{LP-Schouten}
\end{equation}
with
$(\beta_n\contract\nabla)\alpha_m
=\beta_n^{\,r,i_1,\dots, i_{n-1}}
\partial_{r}\alpha_m^{i_n,\dots,i_{n+m-1}}$,
where contraction of lower and upper indexes is always assumed (sum over
$r$ in this case). The key observation now is that, since moments are symmetric,
the antisymmetric part of the tensor expression in square brackets yields
zero contribution. Consequently, one can rewrite the Lie Poisson bracket for
the moments as
\begin{align}
\left\{F,G\right\}&=\,-
\sum_{n,m=0\,}^\infty
%\left\langle
\int\!
A_{m+n-1}\,\,\textit{\large$\hat{S}$}\!\left[ n
\left(\frac{\delta F}{\delta A_n}\contract\,\nabla\right)\frac{\delta G}{\delta A_m}
-
m
\left(\frac{\delta G}{\delta A_m}\contract\,\nabla\right)\frac{\delta F}{\delta A_n}\right]
%\right\rangle
\,{\rm d}^N{\bf q}
\\
&=:\,-
\sum_{n,m=0\,}^\infty
\left\langle
A_{m+n-1},
\left[
\frac{\delta F}{\delta A_n},
\frac{\delta G}{\delta A_m}
\right]
\right\rangle
\,,
\end{align}
where $\hat{S}$ is the symmetrizer operator and the square brackets $[\cdot,\cdot]$ define \cite{Schouten} the {\it Schouten concomitant} (or {\it symmetric Schouten bracket}), 
\rem{ %%%%%%%%%%%%%%%%%%%%%%%%%%%5 BEGIN REM %%%%%%%%%%%%%%%%%%%%%%%%%%%%%%%5
\comment{CT: In order to understand what happens, one writes
\[
[\beta_n,\alpha_n]=\hat{S}[\hat{S}b_n\,,\hat{S}a_n]
%+\hat{A}[\hat{S}b_n\,,\hat{S}a_n]
\]
Thus, pairing with a covariant tensor $T_{m+n-1}$ yields
\[
\langle \hat{S} T_{n+m-1}, [\hat{S}b_n\,,\,\hat{S}a_m]\rangle
=
\langle A_{n+m-1}, [\hat{S}b_n,\hat{S}a_m]\rangle
=
\langle A_{n+m-1}, [\beta_n,\alpha_m]\rangle
\]
}
} %%%%%%%%%%%%% END REM 
which is a well-known object in differential geometry \cite{KoMiSl93}. 
Upon denoting the Schouten
bracket as $[\beta_n,\alpha_m]={\rm ad}_{\beta_n}\alpha_m$, one may introduce the infinitesimal coadjoint action ad$^*$ as its dual,
\begin{equation}\label{co-adj}
\sum_{k,n=0}^\infty\left\langle\, {\rm ad}^*_{\beta_n} \,A_{k\,},\,\alpha_{k-n+1}\right\rangle:=
\sum_{k,n=0}^\infty\left\langle\, A_{k\,},\,{\rm ad}_{\beta_n}\,\alpha_{k-n+1}\right\rangle
\,.
\end{equation}
Thus, ${\rm ad}^*_{\beta_n} \,A_k$ yields a covariant $(k-n+1)$-tensor density. The moment equations may be obtained from (\ref{LP-Schouten}) and written in terms of the Hamiltonian $\mathcal{H}$ as
\begin{equation}\label{mom-eq}
\frac{\partial  A_m}{\partial t}
=\sum_{n=0}^\infty\,\textrm{\large ad}^*_\text{\normalsize$\frac{\delta \mathcal{H}}{\delta\! A_n}$}\, A_{m+n-1}
=
\{\,A_m\,,\,\mathcal{H}\,\}
\,.
\end{equation}
An explicit expression for ad$^*_{\beta_n}A_{m+n-1}$ may be obtained from its definition by a direct calculation as
\begin{equation}
-\,{\rm ad}^*_{\beta_n}A_{m+n-1}=
n\,
(\beta_n\contract\nabla)A_{m+n-1}
+
n\,
(\nabla\contract\beta_n)A_{m+n-1}
+m\,
A_{m+n-1}\contract\nabla\beta_n
\,,
\end{equation}
where we use the notation
\begin{align*}
(\beta_n\contract\nabla)A_{k}
&:=
\beta_n^{\,r,i_1\dots,i_{n-1}\,}\pa_{\,r\,}(A_{k})_{i_1,\dots,i_{k}}
\\
(\nabla\contract\beta_n)A_{k}
&:=
(A_{k})_{i_1,\dots,i_{k}\,}\pa_{\,r\,}\beta_n^{\,r,i_1\dots,i_{n-1}}
\\
A_{k}\contract\nabla\beta_n
&:=
(A_{k})_{\,i_1,\dots,i_{k}\,}\pa_{\,r\,}\beta_n^{\,i_1,\dots,i_n}
\end{align*}
so that the summation is intended over all the repeated indexes. For the
case of Vlasov
dynamics, one has \makebox{$k=m+n-1$} and since $m\geq0$, the contravariant
indexes are all contracted, so that the resulting quantity is always a (symmetric) covariant tensor density.

%\subsection{Discussion}
Thus the tensor interpretation of the
moments provides a direct identification between the moment Lie bracket and the symmetric Schouten bracket (or concomitant). This bracket was known to Schouten as an invariant differential operator (substitution of covariant
derivatives is always possible) and its relation with
the polynomial algebra of the phase-space functions is very well known. 

\subsection{History of the Schouten concomitant}
The symmetric Schouten bracket has been object of some studies in differential geometry \cite{BlAs79,DuMi95,Ta1996,LeOv2000} and has found applications in various aspects of quantum mechanics \cite{BlAs75,BlGh1975}. Also, since Killing tensors on a Riemannian (or pseudo-Riemannian) manifold are symmetric contravariant tensors, they have been associated with the Schouten concomitant in \cite{Benn06,So1973} and it has been pointed out \cite{Ge1970,Benn06} that they can be defined as the moment sub-algebra consisting of symmetric tensor fields commuting with the contravariant metric $g^{-1}$, under the Schouten concomitant.  

To our knowledge, the relation of the Schouten concomitant with Lie-Poisson dynamics for the Vlasov moments is new. However, it is perhaps not entirely unexpected. 
In fact, the Lie-Poisson bracket functional (\ref{LP-Schouten}) was already considered by Kirillov \cite{Ki82} in the context of invariant differential operators.
This is not surprising, since Kirillov was the first to introduce the Lie-Poisson structure for Hamiltonian systems, also known as the Kostant-Kirillov structure. In particular, he noticed how this bracket functional can generate what we have called the coadjoint operator (${\rm ad}^*_{\beta_h}\,A_k$),
``which, apparently, has so far not been considered'', he remarked in 1982.
(The Kupershmidt-Manin operator was known since 1977, although its association with the Schouten concomitant had apparently not been noticed.) What Kirillov considered were the cases $h=1$, which is the Lie derivative, and  $h=k$, which is often called {\it Lagrangian Schouten concomitant}.

\subsection{Specializations of the Schouten concomitant}
As one may verify directly, the Schuouten concomitant ad$_{\beta_1}\alpha_1$ among two vector fields $\beta_1$ and $\alpha_1$ is equal to the Jacobi-Lie bracket $[\beta_1,\alpha_1]_\textit{\tiny\!J\!L}={\rm ad}_{\beta_1}\alpha_1$, which is equal, in turn, \cite{MaRa99} to the Lie derivative
$\pounds_{\beta_1}\alpha_1=[\beta_1,\alpha_1]_\textit{\tiny\!J\!L}$.
One also obtains ad$_{\beta_1}\alpha_n=\pounds_{\beta_1}\alpha_n$ and analogously the dual operation ad$^*$ is (minus) the Lie derivation ad$^*_{\beta_1}A_n=-\,\pounds_{\beta_1}A_n$.
Moreover, it is interesting to notice the chain of relations,
\[
\left\langle{\rm ad}^*_{\beta_n}\,A_n,\,\alpha_1\right\rangle
=
-\left\langle A_n,\,{\rm ad}_{\alpha_1}\,{\beta_n}\right\rangle
=
-\left\langle A_n,\,\pounds_{\alpha_1}\,{\beta_n}\right\rangle
=:
\left\langle A_n\diamond \beta_n,\,{\alpha_1}\right\rangle
\,,
\]
where $\langle\cdot,\,\cdot\rangle$ is the $L^2$ pairing and the {\it diamond\,} operation $\diamond$ denotes (minus) the dual
action of the Lie derivative, following the work of Holm, Marsden and Ratiu \cite{HoMaRa}. Thus, one obtains the following expression for the Lagrangian Schouten concomitant:
\[
{\rm ad}^*_{\beta_n}\,A_n=\beta_n\diamond A_n
\,.
\]
Consequently, the equation for the first-order moment $A_1$ (the fluid
momentum) may be written
from (\ref{mom-eq}) as 
\[
\frac{\pa{A}_1}{\pa t} = \sum_n\beta_n\diamond A_{n\,}
\,.
\]
In fluid dynamics, the first two terms in the sum represent the fluid transport ($n=1$) and the pressure pressure contribution ($n=0$) for fluid motion \cite{HoMaRa}.

\subsection{The graded structure of the moment algebra}
The Lie algebra structure generated by the Schouten concomitant is that of
a graded Lie algebra $\mathfrak{g}=\bigoplus_i\mathfrak{g}_i$. The filtration \makebox{$[\mathfrak{g}_n,\mathfrak{g}_m]=\mathfrak{g}_{m+n-1}$} allows the subspace $\mathfrak{g}_1=\mathfrak{X}$ of vector fields as a particular sub-algebra, whose underlying Lie group consists of the diffeomorphisms on the configuration manifold \cite{GiHoTr2007}. The largest possible sub-algebra corresponds physically \cite{GiHoTr2007} to the barotropic fluid motion and it is given by $\mathfrak{g}_1\oplus\mathfrak{g}_0$, where $\mathfrak{g}_0$ consists of scalar functions. 

\subsection{Momentum map considerations}
The Schouten concomitant establishes an isomorphism $\mathfrak{g}\to\mathcal{F}$ between the moment algebra $\mathfrak{g}$ and the polynomials in the $p$'s, which are embedded in the
space $\mathcal{F}$ of phase-space functions. This map is also a Lie algebra homomorphism and this is a key step in the following discussion, which explains how the operation of taking the moments is a momentum map arising from a Lie algebra action on the Vlasov distribution. Indeed we shall see that the momentum map is
exactly the dual of this Lie algebra homomorphism. 

Analogous considerations \cite{HoLySc1990} on the structure of the moment Lie algebra also hold for the Lie algebra of statistical moments, which are more simply defined as symmetric tensors, rather than tensor fields. 
%This difference is important in the search for moment invariants, although
%we leave this topic for future developments.

\section{A momentum map for kinetic moments}
\label{mom map-sec}

As a starting point, one recalls \cite{MaRa99} the definition of a momentum
map: Given a Poisson
manifold (a manifold $P$ with a Poisson bracket 
$\{\cdot,\cdot\}$ defined on the smooth functions $\mathcal{F}(P)$) 
and a Lie group $G$ acting on it by Poisson maps (i.e. the Poisson bracket is preserved), a momentum map is defined as a map $\mathbf{J}:P\rightarrow\mathfrak{L}^*$ such that
\[
\{F(p),\langle \mathbf{J}(p),\xi\rangle\}=\xi_P[F(p)]
\qquad\quad\forall\, F\in\mathcal{F}(P),\quad\forall\,\xi\in\mathfrak{L}
\,,
\]
where $\mathfrak{L}$ is the Lie algebra of $G$, the notation $\langle\cdot,\,\cdot\rangle$
indicates the pairing between $\mathfrak{L}^*$ and $\mathfrak{L}$, and $\xi_P$ is the vector field given by the infinitesimal generator
\[
\xi_P\,(p)=\left.\frac{d}{dt}\right\vert_{t=0}\!\!\!e^{t\xi}\cdot p \qquad \forall\,p\in
P
\,.
\]
Thus, whenever the infinitesimal generator is known, a corresponding momentum
map can be defined. Indeed, the Lie group action is not strictly necessary
to this definition. Rather what is necessary is the {\it Lie algebra action}
\cite{MaRa99}, i.e. the map associating the infinitesimal generator $\xi_P$
to the generic Lie algebra element $\xi$.

In the case of moments, the Lie-Poisson structure (\ref{KMLP}) involves a
Lie algebra which is defined by the term in square brackets, while the moments
$A_n$ belong to the dual Lie algebra. Moreover, moments are linear functionals of the Vlasov distribution, $f$, as in (\ref{mom-def}). The aim of this section is the show that the operation of taking the moments is not only a Poisson map (Gibbons' theorem \cite{Gi1981}), but is also a momentum map, arising
from a Lie algebra action on the phase space density $f$. Therefore, the Poisson manifold $P$ in the definition
of $\bf J$ is the space of densities $P=\mathcal{F}^*(T^*Q)$, while
the moment Lie algebra is defined through the expression in (\ref{KMLP}).
Consequently, a single element $\xi\in\mathfrak{L}$ is identified with a whole sequence
$\beta_n$ and the corresponding infinitesimal generator $\xi_P$ is denoted by $\beta_P$. In the remainder of this section the bracket $\{\cdot,\,\cdot\}$ always denotes the canonical Poisson bracket, while $\{\cdot,\,\cdot\}_V$ stands for the Vlasov bracket (\ref{VLP}).

\subsection{The Lie algebra action and its momentum map}
In order to show that moments constitute a momentum map, one starts with
the Vlasov equation written in the advective form
\[
\frac{\pa f}{\pa t}+\textit{\large\pounds}_{{\bf X}_H}\,f=0
\,,
\]
where the phase-space function $H$ is the single-particle Hamiltonian, which
is written from the Vlasov Lie-Poisson structure as $H=\delta\mathcal{H}/\delta f$. Substitution of expression (\ref{chain}) then yields the Vlasov equation in the form
\begin{equation}
\frac{\pa f}{\pa t}+
%\textit{\large\pounds}_{{\bf X}_\text{\footnotesize$p^n\beta_n(q)$}}\,f=0
\bigg\{f\,,\,\sum_n\, p^{n\,}\frac{\delta\mathcal{H}}{\delta A_n}\bigg\}=0
%\qquad\hbox{(no sum on $n$)}
\,.
\end{equation}
The implied summation on repeated indices will be omitted from now on. We will also use the notation for the one-dimensional
case, since the extension to higher dimensions may be madel by following the arguments in the previous section. The bracket term is evidently determined
by the Lie derivative $\pounds_{\bf X}\,f$ along the Hamiltonian vector field
${\bf X}$ defined through the fiberwise polynomial $\,p^n\,{\delta\mathcal{H}}/{\delta A_n}\,$. That is, the bracket term is determined by the symplectic Lie algebra action on its dual. However, by the usual isomorphism between polynomials and symmetric tensors, this term may also be considered as arising from the
{\it moment} Lie algebra action on the phase space densities, so that
\begin{equation}\label{LA-action}
\beta\,\cdot f=-\,\left\{f,\,p^n\,\beta_n\right\}
\,,
\end{equation}
where $\beta$ denotes here a whole sequence of symmetric contravariant tensor
fields $\beta:=\left\{\beta_n\right\}_{n\,\in\,\mathbb{N}}$.
 
At this point, one defines the map ${\bf J}:\mathcal{F}^*(T^*Q)\to\mathfrak{g}^*$
($\mathfrak{g}$ is the Lie algebra of contravariant tensor
fields, as above) from Vlasov distributions to moments as in (\ref{mom-def})
(or as in (\ref{mom3D}), when including the basis). A direct verification
shows that this is a momentum map. Namely,
\begin{align*}
\big\{F(f),\langle \mathbf{J}(f),\beta\,\rangle\big\}_{\!V}
&=
%\left\langle
\int\hspace{-3mm}\int\!
f(q,p)\left\{\frac{\delta F}{\delta f}, \frac{\delta}{\delta f}\!\int\!\left(\int p^n\,f(q,p)\,{\rm d}p\right)\beta_n(q)\,{\rm
d}q\right\}
%\right\rangle
{\rm d}q\,{\rm d}p
\\
&=
\int\hspace{-3mm}\int\!
f(q,p)\left\{\frac{\delta F}{\delta f},\, p^n\,\beta_n(q)\right\}
%\right\rangle
{\rm d}q\,{\rm d}p
\\
&=-\!
\int\hspace{-3mm}\int
\Big\{f(q,p),\, p^n\,\beta_n(q)\Big\}\,
\frac{\delta F}{\delta f}
%\right\rangle
\,\,{\rm d}q\,{\rm d}p
\,=\,\beta_{P}[F(f)]
\,,
\end{align*}
where the last line is justified by integration by parts and by the definition of the moment Lie algebra action in (\ref{LA-action}) (recall that $P=\mathcal{F}^*(T^*Q)$ as above).

\subsection{Infinitesimal equivariance}

This section shows that the momentum map ${\bf J}(f)$ defining the moments is (infinitesimally) {\it equivariant} \cite{MaRa99}, i.e., 
\begin{equation}\label{equivar}
\beta_P\left[\left\langle{\bf J}(f),\,\gamma\right\rangle\right]=
\left\langle{\rm ad}^*_{\beta\,}\,{\bf J}(f),\,\gamma\right\rangle
\qquad\quad
\forall\,\,\beta,\gamma\in\mathfrak{g}
\qquad
\forall\,\,f\in P=\mathcal{F}^*(T^*Q)
\,,
\end{equation}
where the coadjoint operator ${\rm ad}^*$ is defined as in (\ref{co-adj}).
Although one can easily verify this property by direct substitution upon
following the calculations in \cite{Gi1981,GiHoTr05,GiHoTr2007}, we prefer to show equivariance through the identification
of $\bf J$ with the dual $\alpha^*$ of a Lie algebra homomorphism $\alpha$,
which is known \cite{MaRa99} to be an equivariant momentum map in the general case. This is preferable, since it enriches the geometric character of $\bf J$.

A {\it Lie algebra homomorphism} is a map $\alpha:\mathfrak{w}\to\mathfrak{h}$ between two Lie algebras $\mathfrak{w}$ and $\mathfrak{h}$ satisfying the following property
\[
\alpha\left(\left[\xi,\eta\right]_\mathfrak{w}\right)=\left[\alpha(\xi),\alpha(\eta)\right]_\mathfrak{h}
\,.
\]
In the present case, one has $\mathfrak{w}=\mathfrak{g}$ (the moment algebra) and $\mathfrak{h}=\mathcal{F}(T^*Q)$.
The homomorphism $\alpha$ is actually provided by the usual isomorphism
between symmetric tensors and polynomials, so that \makebox{$\alpha:\beta\mapsto p^n\beta_n$}. The Lie brackets on $\mathfrak{g}$ and $\mathcal{F}$ are defined
by the Schouten concomitant and the canonical Poisson bracket respectively.
A trivial calculation shows that
\[
\alpha\big(\!\left[\beta,\gamma\right]\!\big)
=p^{m+n-1}\left[\beta_n,\gamma_m\right]
=
\left\{p^n\beta_{n\,},\,p^m\gamma_m\right\}
=\left\{\alpha(\beta),\alpha(\gamma)\right\}
\,.
\]
Consequently, 
\[
\big\langle{\bf J}(f),\beta\big\rangle_{\mathfrak{g}}
=
\int\!  A_n(f)\,\beta_n\,{\rm d}q
\,
=\!\int\hspace{-3mm}\int\!
f\, p^n\beta_n\,
\,{\rm d}q\,{\rm d}p
=\big\langle f, \alpha(\beta)\big\rangle_{\!\mathcal{F}}
\,,
\]
so that ${\bf J}=\alpha^*$. 
The (infinitesimal) equivariance (\ref{equivar}) of $\bf J$ arises naturally,
since it is a general fact \cite{MaRa99} that any dual Lie algebra homomorphism is an equivariant momentum map. Notice that from (\ref{equivar}) one can characterize equivariance also by
\[
\big\{\!
\left\langle
{\bf J}(f),\beta
\right\rangle
,\,
\left\langle
{\bf J}(f),\gamma
\right\rangle\!
\big\}_{\!V}
=
\big\langle
{\bf J}(f),\,\left[\beta,\,\gamma\right]
\big\rangle
\,,
\]
which is verified by direct substitution.
At this point, Gibbons' theorem on Vlasov kinetic moments becomes justified by the equivariance of the momentum map
${\bf J}(f)=\left\{A_n\right\}_{n\,\in\,\mathbb{N}\,}$.
Indeed, any equivariant momentum map is Poisson \cite{MaRa99}.

The term ``infinitesimal'' used above refers to the fact that the momentum
map is {\it not} known to arise from a group action. Indeed, the group action
associated to moment dynamics is still unknown and we leave this issue for future work. In particular, one would like to identify the moment algebra
$\mathfrak{g}$ as the tangent space at the identity $T_e G$ of some Lie group
$G$, whose infinitesimal action is given by (\ref{LA-action}).

\section{A bosonic Fock space for the moments}
\label{Fock-sec}

\subsection{Bosonic structure of the moment Fock space}
It is important to notice that if one fixes a point $q$ on the base manifold $Q$, then the $\beta$'s can be considered as simple tensors belonging to $\bigvee T_q Q$, the symmetric algebra on the tangent fiber $T_q Q$. Here the notation $\bigvee W$ stands for $\bigoplus_{n}\left(\bigvee^{\,n} W\right)$, where $\bigvee$ denotes the symmetric tensor product \cite{Shaw}, so that $\bigvee^{\,n} W:=\hat{S}\left(\bigotimes^n W\right)$ and $\hat{S}$ is the symmetrizing operator. Thus, the space of moments can be identified with the {\it Fock space} on the tangent fiber $T_q Q$ corresponding to the Bose-Einstein statistics \cite{Fock}.
\begin{remark}[The bundle of Fock spaces]
In more generality, the moment algebra is composed 
%of sequences $\beta=\left\{\beta_n\right\}_{n\in\mathbb{N}}$
of tensor fields, mapping the configuration manifold $Q$ to the bundle of Fock spaces
\[
STQ\,:=\bigcup_{q\in Q}\left(\bigvee T_q Q\right)
\,.
\]
The geometric properties of such bundle are not completely understood. An important question would concern its dual $ST^*Q\,:=\bigcup_{q\!}\left(\bigvee T_q^* Q\right)$. Indeed, it is well known that the cotangent bundle
$T^*Q=\bigcup_{q\!}T^*_q Q$ is characterized by the symplectic form $\omega={\rm
d}\theta={\rm
d}(p\,{\rm d}q)$ and it is not clear what role (if any) this symplectic form may play on $ST^*Q$. Such questions will be addressed in future work.
\end{remark}

\subsection{Creation and annihilation operators}
In second quantisation one also defines creation and annihilation operators. Within the moment Lie algebra, one may choose a vector field $\alpha_1$ and use it to construct the {\it creation operator} acting on a symmetric $n$-tensor
field $\beta_n$ as
\begin{equation}
a(\alpha_1)\beta_n:=\alpha_1 \vee \beta_n:=\hat{S}\left(\alpha_1 \otimes \beta_n\right)
\,.
\end{equation}
The {\it annihilation operator} can be defined as the dual
\begin{equation}
\big\langle
A_{n+1},\,
a(\alpha_1)\beta_n
\big\rangle
=
\big\langle
\big(\alpha_1\contract \hat{S}A_{n+1}\big),\,
\beta_n
\big\rangle
=:
\big\langle
a^*(\alpha_1)A_{n+1},\,
\beta_n
\big\rangle
\,,
\end{equation}
so that 
\begin{equation}
a^*(\alpha_1)A_{n}=\left(\alpha_1\contract A_{n}\right)
\,,
\end{equation}
where the contraction is consistently symmetrized. Consequently these operators are such that \makebox{$a:T_q Q\times\bigvee^n T_q Q\to \bigvee^{n+1}T_q Q\,$} and
\makebox{$a^*:T_q Q\times\bigvee^n T_q^* Q\to \bigvee^{n-1}T_q^* Q$}. It is obvious that these operators commute
\begin{equation}
\left[a(\alpha_1),\,a(\gamma_1)\right]=0=\left[a^*(\alpha_1),\,a^*(\gamma_1)\right]
\,.
\end{equation}
The fact that $a$ acts on symmetric powers
of the tangent space is not essential: one only needs to think of $a$ and
$a^*$ as operations $a:W\times\bigvee^n W\to \bigvee^{n+1}W$ and
\makebox{$a^*:W\times\bigvee^n W^*\to \bigvee^{n-1}W$} \cite{Shaw}, whose properties
are unchanged when $W$ can be identified $W^*$. In particular,
if one considers the moment tensor-densities $A_n$ of the form 
$A_n=\Pi_n\otimes A_0$ (with $\Pi_0\!:=\!1$), one can construct the operation \begin{equation}
a(\Theta_1)A_n:=\Theta_1\vee A_n:=\left(\Theta_1\vee\Pi_n\right)\otimes A_0
\end{equation}
for any one-form $\Theta_1\in T_q^* Q$ (one has $a(\Theta_1)A_0:=\Theta_1\otimes A_0$). By analogy, one writes the dual
\begin{equation}
a^*(\Theta_1)\beta_n=\left(\Theta_1\contract\beta_n\right)
\,.
%\otimes A_0
\end{equation}
By using these definitions one checks that
\[
\left[a^*\left(\alpha_1\right),\,a\left(\Theta_1\right)\right]
=\left(\alpha_1\contract \Theta_1\right){\bf I}
\,.
\]
If one takes $\alpha_1=\pa_{\!q^\textit{\!\scriptsize h}}$ and $\Theta_1={\rm d}q^k$, then
\[
\left[a^*_h,a_k\right]
=\delta_{hk}
\,,
\]
where one defines $a_k:=a\left({\rm d}q^k\right)$ and 
$a^*_h:=a^{*\!}\big(\pa_{\!q^\textit{\!\scriptsize h}}\big)$. These commutation relations
are identical to those used in second quantization.

%\begin{framed}
It is important to notice that any one form $\Theta_1$ can be identified with a first-order kinetic moment $B_1=\Theta_1\otimes B_0$ arising from
a Vlasov distribution. In more generality, any element $\Theta_n\in\bigvee^n T^*_q
Q$ can be identified with a $n$-th order moment $B_n=\Theta_n\otimes B_0
%\in\bigvee^n T^*_q Q\otimes \mathcal{F}^*(Q)
$,
so that the operators $a(\Theta_1)A_n$ and $a^*(\Theta_1)\beta_1$ can be
considered as operators such that
\[
a(B_1)A_n:=\left(\frac{B_1}{B_0}\vee A_n\right)\otimes A_0
\qquad\qquad
a^*(B_1)\beta_1=\frac{B_1}{B_0}\contract_{\,} \beta_n
\]
At this point, creation and annihilation operators involve only moments $A_n,B_n$ and
their dual variables $\alpha_n,\beta_n$.
%\end{framed}

It may be useful to point out that the physics literature on quantum mechanics
\cite{FeWa03,Fock}
uses the opposite notation for the creation and annihilation operators, so
that they become switched $a\leftrightarrow a^*$. In this paper we adopt
the more mathematical convention in \cite{Shaw} and no confusion should arise
from this choice.

\subsection{Cold plasma solution from creation operators}
A particular case where the creation operator can be applied is the {\it
cold plasma solution} of the Vlasov equation, which is written as 
$f(q,p,t)=A_0(q,t)\,\delta(p-\Pi(q,t))$. This yields a moment hierarchy such that
\[
A_n=A_0\,\Pi^n
=A_0\left(\frac{A_1}{A_0}\right)^{\!n}
\,.\]
It is now clear that this moment hierarchy may be obtained by iterating the composition of the creation operator
\[
A_n=\left[a\left(A_1\right)\right]^{n}A_0
%=a^{n\!}\!\left(\frac{A_1}{A_0}\right)A_0
\,.
\]
%where $a\left(\Pi\right)A_0=\Pi\otimes A_0=A_1$.
Thus, this case provides a useful interpretation of creation and annihilation operators in the moment Fock space. Any moment is constructed by recursively
applying the creation operator on the {\it vacuum state} $A_0$, so that $A_1=a(A_1)A_0=
\Pi\otimes A_0$.

By analogy, for a vector field $\alpha_1$ we can see that
\[
a^*(\alpha_1)A_n=\left(\alpha_1\contract\Pi^n\right)\otimes A_0=\left(\alpha_1\contract\Pi\right) A_{n-1}
\]
and, thus,
\[
\left[a^{*\!}\!\left(\Pi^\sharp\right)\right]^{n\!} A_n=|\Pi|^{2n}A_0
\,,
\]
where the sharp $\sharp$ indicates the usual operation of raising indexes
and $|\Pi|^2:=\Pi\contract\Pi^\sharp$.

\subsection{Occupation numbers}

The analogy with the bosonic Fock space in second quantisation proceeds further by noticing that the multi-index notation introduced in equation (\ref{mom-def3D}) corresponds to the treatment of {\it occupation numbers} $\sigma_1,\dots,\sigma_N$,
for the symmetric tensors $A_n(q)$ (cf. e.g. \cite{Shaw}). Thus, the interpretation in terms of symmetric tensor fields is also equivalent to the geometric characterization of the moments $A_\sigma$. The operators $a$ and $a^*$ also have an equivalent in
terms of occupation numbers. For example, take a Lie algebra variable 
$\beta_\sigma=\beta_{\sigma_1,\dots,\sigma_N}$ with $\sum \sigma_i=n$ and a vector field $\alpha_{1_\textit{\scriptsize k}}$. The creation operator becomes
\[
a\left(\alpha_{1_\textit{\scriptsize k}}\right)\beta_\sigma=
\big(\alpha_1\vee\beta_n\big)_{\sigma_1,\dots,\sigma_k+1,\dots,\sigma_N}
\,,
\]
so that the $k$-th occupation number $\sigma_k$ is increased by 1. In particular, one may focus on the basis and introduce the ket notation
\[
\left|\sigma_1,\dots,\sigma_N\right\rangle:=\left(\pa_{q^1}\right)^{\sigma_1}\dots\left(\pa_{q^N}\right)^{\sigma_N}
=
\left|\sigma_1\right\rangle\dots\left|\sigma_N\right\rangle
\,,
\]
so that
\begin{align}
a_h\left|\sigma_1,\dots,\sigma_N\right\rangle&=
a\big(\pa_{\!q^\textit{\!\scriptsize h}\,}\big)
\left|\sigma_1,\dots,\sigma_N\right\rangle
=\sqrt{\sigma_h+1\,}\left|\sigma_1,\dots,\sigma_h+1,\dots,\sigma_N\right\rangle
\,,
\\
a_k^*\left|\sigma_1,\dots,\sigma_N\right\rangle&=
a^{*\!}\big({\rm d}q^k\big)\left|\sigma_1,\dots,\sigma_N\right\rangle
=\sqrt{\sigma_{k\,}}\left|\sigma_1,\dots,\sigma_k-1,\dots,\sigma_N\right\rangle
\,.
\end{align}
Indeed, it is well known in quantum theory \cite{FeWa03} that all the quantum-mechanical properties of these operators follow {\it uniquely} from the commutation relations in the previous section.
By proceeding analogously, one defines the bra corresponding to the moment variable as
\[
\left\langle\sigma_1,\dots,\sigma_N\right|:=\left({\rm d}{q^1}\right)^{\sigma_1}\dots\left({\rm d}{q^N}\right)^{\sigma_N}\otimes A_{0\,}{\rm d}Vol
\,.
\]
(The ``vacuum density'' $A_0$ is not a problem since it yields 1 under
the $L^2$ pairing.) One then writes
\begin{align}
\left\langle\sigma_1,\dots,\sigma_N\right|a_k&=
\left\langle\sigma_1,\dots,\sigma_N\right|a\big({\rm d}q^k\big)
=
\sqrt{\sigma_k+1\,}\left|\sigma_1,\dots,\sigma_k+1,\dots,\sigma_N\right\rangle
\,,
\\
\left\langle\sigma_1,\dots,\sigma_N\right|a^*_h&=
\left\langle\sigma_1,\dots,\sigma_N\right|a^{*\!}\big(\pa_{\!q^\textit{\!\scriptsize h}}\big)
=
\sqrt{\sigma_k\,}\left|\sigma_1,\dots,\sigma_h-1,\dots,\sigma_N\right\rangle
\,.
\end{align}
Thus, the moment framework on the bosonic Fock
space is completely consistent with the standard treatment of second quantization.
For example, one can also define the {\it number operator} $\mathcal{N}:=\sum_k
a_k a^*_k$ so that
\begin{align}
\mathcal{N}\left|\sigma_h\right\rangle=
\sqrt{\sigma_{h}}\,\,\sum_k  a_k \left|\sigma_h-1\right\rangle
=
\sigma_h\left|\sigma_h\right\rangle
\end{align}
and more particularly,
\begin{align}
\mathcal{N}\left|\sigma_1,\dots,\sigma_N\right\rangle=\sum_{k=1}^N\,\sigma_k=n
\end{align}
so that the operator $\mathcal{N}$ corresponds to the moment order, instead
of the number of particles occupying a certain quantum system.
%\comment{CT: normalization factors should appear! Define the occupation
%number operator $N_k=a_{k\,} a_k^*$!} 
%Analogously, the annihilation $a^*(\alpha_{1_\textit{\scriptsize k}})$ decreases
%$\sigma_k$ by 1 when applied on $A_\sigma$.

\rem{ %%%%%%%%%%%%%%%%%%%%%% BEGIN REM %%%%%%%%%%%%%%%%%%%%%%%
All the arguments above hold for a fixed point $q\in Q$ in the configuration
space. However they can be extended to the whole $Q$ as follows. 
For $n\in\mathbb{N}$ fixed, one has 
\[
\beta_n:Q\to S^n TQ\,:=\bigcup_{q\in Q\!}\left(\bigvee\,\!\!^{n\,} T_q Q\right)
\]
which consistently reduces to the ordinary definition of vector fields, when
$n=1$ so that \makebox{$\beta_1:Q\to TQ$}, where $TQ$ is the tangent bundle of $Q$. 
One can proceed further by defining moments as dual to the $\beta$'s. To
this purpose one introduces the dual Fock space $\bigvee T_q^*Q$. However,
since kinetic moments are tensor-densities, one has to construct the direct
product space $\left(\bigvee T_q^*Q\right)\bigotimes\mathcal{F}^*(Q)$. At
this point, the moments are maps such that
\[
A_n:Q\to\bigcup_{q\in Q\!}\left(\left(\bigvee\,\!\!^{n\,} T_q^* Q\right)\bigotimes
\mathcal{F}^*(Q)\right)
\]
and one can also define the dual bundle
\[
S^*Q\,:=\bigcup_{q\in Q\!}\left(\left(\bigvee T_q^* Q\right)\bigotimes
\mathcal{F}^*(Q)\right)
\]
} %%%%%%%%%%%%%%%%%%%%%%%%%% END REM %%%%%%%%%%%%%%%%%%%%%%%%%%%%%%

\begin{remark}[Statistical moments] Since the treatment of statistical moments is analogous to that of kinetic moments and it involves the symmetric tensor algebra in a similar fashion, we expect no obstacles in transferring the same Fock space treatment to statistical moments.
\end{remark}

\rem{ %%%%%%%%%%%%%%%%%%%% BEGIN REM %%%%%%%%%%%%%%%%%%%%%%%%%%%%%%%%%%%
\begin{remark} [The Fermi-Dirac statistics] The moment Fock space above involves the bosonic statistics
because of the symmetric nature of the Schouten concomitant. However there exist antisymmetric analogues whose most famous example is probably the Scouten-Nijenhuis
bracket (cf. e.g. \cite{MaRa99}), which is mainly used in Poisson geometry. It would be interesting to investigate
the possibility of a fermionic Fock space for the Scouten-Nijenhuis bracket.
\end{remark}
} %%%%%%%%%%%%%%%%%%%%%%%%%% END REM %%%%%%%%%%%%%%%%%%%%%%%%%%%%%%

\section{Extension to anisotropic interactions}
\label{anisotropy-sec}

The Vlasov kinetic equation (\ref{Vlasov-eq}) regulates multi-particle dynamics
in phase space of point-like particles. What happens if the particles have
a shape in physical space (e.g., they are rigid bodies)? This section addresses the question of how to generalize the geometric moment
framework to particles undergoing anisotropic interactions. We base our analysis on a particular kind of Vlasov equation, which was proposed by Gibbons, Holm and Kupershmidt (GHK) \cite{GiHoKu1982,GiHoKu1983} for obtaining the equations
of chromohydrodynamics of quark-gluon plasmas. The GHK approach defines a density variable which depends not only on phase space coordinates, but also on an extra degree
of freedom $\bg$ (e.g., orientation). In order to respect the geometric structure
of the Vlasov equation, this variable is supposed to belong to a (dual) finite-dimensional
Lie algebra $\mathfrak{h}$, so that the GHK-Vlasov equation is written as
\begin{equation}
\frac{\partial f}{\partial t}+
\Big\{f,H\Big\}\,+\,\left\langle{\bg},\left[\frac{\partial f}{\partial {\bf \bg}},\frac{\partial H}{\partial {\bf \bg}}\right]_{\!\mathfrak{h}}\right\rangle=0
\,.
\end{equation}
At this point, one may introduce two possible types of moments:
the {\it Smoluchowski moments} (placed in this framework by Holm, Putkaradze and Tronci \cite{HoPuTr2007a,HoPuTr2008})
\[
A_n(q,\bg,t)\,:=\int p^n\,f\,\,{\rm d}^N p
\]
and the {\it GHK-moments} \cite{GiHoKu1982,GiHoKu1983}
\begin{equation}
A_{n,\nu}(q,t)\,:=\int p^n\,\bg^\nu \,f\,\,{\rm d}^N p\,\,{\rm d}^{K\!}\bg
\,,
\end{equation}
where $n,\nu\in\mathbb{N}$ and $K={\rm dim}\,\mathfrak{h}$. Thus $\bg^\nu$
is again a tensor $\nu$-th power of $\bg$. 
The approach in the previous sections can be generalized to both of these
kinds of moments. However, since the GHK-moments involve also powers of $\bg$
they appear to be geometrically more interesting and we choose to focus on them for the remainder of this section.

In order to generalize the Schouten bracket, one starts with the Lie-Poisson
structure
\begin{equation}\label{VLP-GHK}
\left\{F,G\right\}[f]\,=\int\hspace{-3mm}\int\hspace{-3mm}\int \!f(q,p,\bg,t)\,\left(\left\{\frac{\delta F}{\delta f},\frac{\delta
G}{\delta f}\right\}+\left\langle\bg,\left[\frac{\partial}{\partial {\bf \bg}}\frac{\delta F}{\delta f},\frac{\partial}{\partial {\bf \bg}}\frac{\delta G}{\delta f}\right]_{\!\mathfrak{h}}\right\rangle\right)\,{\rm d}q\,{\rm d}p\,{\rm d}^{K\!}\bg
\,.
\end{equation}
The chain rule allows one to write
\begin{equation}\label{GHK-chain}
\frac{\delta F}{\delta f}=\sum_{n,\nu}\,p^n\bg^\nu\!\contract\,\frac{\delta F}{\delta A_{n,\nu}}
\end{equation}
and a direct substitution into equation (\ref{VLP-GHK}) yields \cite{Tronci}
\begin{equation}\label{GHK-Schouten}
\left\{F,G\right\}\,
=
\bigg\langle
A_{\,m+n-1,\,\nu+\mu\,},
\Big[\alpha_{m,\nu},\,\beta_{n,\mu}\Big]
\bigg\rangle
+
\Big\langle
A_{\,m+n,\,\nu+\mu-1}\,,\,
\Big[\alpha_{m,\nu},\,\beta_{n,\mu}\Big]_C
%C\contract \,\alpha_{m,h}\otimes\beta_{n,k}
\Big\rangle
\,,
\end{equation}
where $[\cdot,\cdot]$ is the ordinary Schouten bracket, while $\left[\cdot,\cdot\right]_C$ is the Lie bracket defined
by
\begin{equation}
\left[\alpha_\nu,\beta_\mu\right]_C:=\nu_{\,}\mu_{\,}\, C\contract\left(\alpha_{\nu}\otimes\beta_{\mu}\right)
\,,
\end{equation}
where $\otimes$ denotes the ordinary tensor product (not symmetrized) and $C$ is the structure tensor of $\mathfrak{h}$ with components $C^a_{bc}$.
The whole bracket tensor $\left[\cdot,\cdot\right]_C$ is then identified with its symmetric part by the usual argument. The derivation of (\ref{GHK-Schouten})
is presented in appendix \ref{appendix}.

Thus, the moment dynamics is extended to anisotropic interactions.  It is straightforward to verify that truncating the hierarchy to consider only $A_{0,0},A_{0,1},A_{1,0}$
yields the equations of chromohydrodynamics in \cite{GiHoKu1982,GiHoKu1983}.
Moreover, it is easy to see that the geometric considerations in the previous
sections can be suitably generalized to the anisotropic case. In particular,
moments are now defined as tensors
\[
A_{n,\nu}=\left(A_{n,\nu}\right)_{i_1,\dots,i_n,\,a_1,\dots,a_\nu}
\,{\rm d}q^{i_1}\otimes\dots\otimes{\rm d}q^{i_n}\otimes{\bf e}^{a_1\!}\otimes\dots\otimes{\bf e}^{a_\nu}\otimes{\rm d}^3{\bf q}
\,,
\]
where $\left\{{\bf e}^{1},\dots,{\bf e}^{K}\right\}$ is a basis of $\mathfrak{h}^*$.
Thus these moments are defined on the product of Fock spaces $\bigvee T_q^* Q_{\,}\bigotimes\bigvee_{\!}\mathfrak{h}^*$ and all the second quantisation
techniques can be as well applied to this case. A natural case for future study could be a system of rigid-body particles. In
this case one identifies $\mathfrak{a=so}(3)\simeq\mathbb{R}^3$
and the structure constants are given by the Levi-Civita symbol $\varepsilon_{ijk}$
corresponding to the vector product in $\mathbb{R}^3$. We leave the geometric
investigation of this structure as a possibility for future work.

\section{Conclusions and open questions}
\label{conclusion-sec}

In this paper, the Schouten symmetric bracket has emerged for the first time as a basic object for the applications of Vlasov kinetic theory. In its previous treatments, the Schouten bracket has been appreciated as an interesting invariant differential operator, irrespective of its potential applications in physical problems. However, many potential applications for it may now arise in phase space dynamics, whose primary physical examples are kinetic equations.
The key to seeing the importance of the Schouten bracket for these applications is the identification of kinetic Vlasov moments with tensor densities, rather than simple covariant tensors as in the previous physical literature \cite{GoZhSa80}.

Directions for further investigation of the Schouten bracket naturally arise from the results in this paper. For example, the identification of  the moments with the bosonic Fock space shows how the cold plasma closure is constructed using standard methods in second quantization, in which the vacuum state is given by the particle density $A_0$. In more generality, a bundle of Fock spaces can be defined in order to treat moments as tensor fields, rather than tensors.
The investigation of the geometry of this bundle is another open question that deserves to further investigation. The most remarkable result of the Fock space treatment is the interpretation that it reveals of Kupershmidt's multi-indices as occupation numbers. This natural  interpretation unifies the two approaches and places them into a single geometric framework.

The present geometric treatment also characterizes the moment hierarchy as a momentum map. This property was not entirely unexpected; but it is welcome, because the Lie algebra action (\ref{LA-action}) can now be uniquely defined as an action of symmetric contravariant tensor fields, so that the momentum map is a dual Lie algebra
homomorphism. In this setting, it is clear that the Poisson structure for the moments arises from infinitesimal equivariance of the momentum map. The outstanding question that remains concerns the characterization of the Lie group corresponding to the moment Lie algebra. If this were possible, one could characterize moment dynamics as a form of coadjoint motion \cite{MaRa99}
\[
A(t)={\rm Ad}^*_{\exp\left(t\,\frac{\delta \mathcal{H}}{\delta_{\!} A}\right)}\,A(0)
\,,
\]
where Ad$^*$ indicates the coadjoint group action and $A$ denotes the entire sequence \makebox{$A:=\{A_n\}_{n\in\mathbb{N}}$}. This point is related to the question of global equivariance \cite{MaRa99} and would be crucial for the identification of families of moment invariants analogous to the Poincar\'e invariants of the particle phase space. This identification is known for the statistical moments \cite{HoLySc1990}, but it remains to be discovered for the kinetic moments.

Finally, the geometric framework for the moments has been extended here to apply in the case of anisotropic particle interactions.
Interest in this extension arose from the previous use of kinetic moments in problems of quark-gluon plasma dynamics \cite{GiHoKu1982,GiHoKu1983}.
Additional interest has also arisen recently in physical problems involving the self-assembly of oriented nano-rods \cite{HoPuTr2007a,HoPuTr2008}. This extension of moments to the kinetic theory of particles undergoing anisotropic interactions may also be useful in problems involving the dynamics of complex fluids possessing anisotropic order parameters \cite{Ho2002}.

\section*{Acknowledgements} 
The authors acknowledge B. A. Kupershmidt for helpful comments. 
The work of DDH was partially supported  by the US Department of Energy, Office of Science, Applied Mathematical Research, and the Royal Society of London Wolfson Research Merit Award. The authors also acknowledge the European Science Foundation for partial support through the MISGAM program.

\appendix

\section{Evaluation of the moment brackets}\label{appendix}
This appendix presents the evaluation of the moment Lie-Poisson bracket in the anisotropic case (\ref{GHK-Schouten}). This proceeds by a systematic full calculation involving tensor indexes \cite{Tronci}. The first term in (\ref{GHK-Schouten})
is identical to the ordinary moment bracket (\ref{LP-Schouten}), which is
thus also evaluated in this appendix. One takes the expression in (\ref{VLP-GHK})
and substitutes (\ref{GHK-chain}) with $\delta F/\delta A_{m,\nu}=\alpha_{m,\nu}$
and $\delta G/\delta A_{n,\mu}=\beta_{n,\mu\,}$. At this point one follows the steps below
\begin{align*}
\{\,G\,,\,H\,\}
=&
%\sum_{m,n=0}^\infty
\int\hspace{-3mm}\int\hspace{-3mm}\int \!f \,
\Big\{\alpha_{m,\nu}({\bf q})\contract{\bf p}^m\otimes{\bg}^{\nu},\,\,\beta_{n,\mu}({\bf
q})\contract{\bf p}^n\otimes{\bg}^{\mu}\Big\}_{\!1}
\,{\rm d}^N{\bf q}\wedge{\rm d}^N{\bf p}\wedge{\rm d}^K{\bg}
\\
=&
\int\hspace{-3mm}\int\hspace{-3mm}\int\! f\, \bg^{\mu+\nu\!} \,
\bigg(\,p_{i_1}\dots p_{i_m}\frac{\pa \left(\alpha_m\right)^{i_1,\dots,i_m}}{\pa q^k}\frac{\pa\, p_{j_1\!}\dots p_{j_n}}{\pa p_k}\,\left(\beta_n\right)^{j_1,\dots,j_n}
\\
&\hspace{2.37cm}-
p_{j_1}\dots p_{j_n}\frac{\pa \left(\beta_n\right)^{j_1,\dots,j_n}}{\pa q^h}\frac{\pa\, p_{i_1\!}\dots p_{i_m}}{\pa p_{h}}\,\left(\alpha_m\right)^{i_1,\dots,i_m}
\bigg)
\,{\rm d}^N{\bf q}\wedge{\rm d}^N{\bf p}\wedge{\rm d}^K{\bg}
\\
&+\!
\int\hspace{-3mm}\int\hspace{-3mm}\int\!
f\left\langle\bg,\left[\frac{\pa}{\pa\bg}\Big(\alpha_{m,\nu}({\bf q})\contract{\bf p}^m\otimes{\bg}^{\nu}\Big),\,\frac{\pa}{\pa\bg}\Big(\beta_{n,\mu}({\bf q})\contract{\bf p}^n\otimes{\bg}^{\mu}\Big)\right]\right\rangle
{\rm d}^N{\bf q}\wedge{\rm d}^N{\bf p}\wedge{\rm d}^K{\bg}
\\
=&\,
%\bigg\langle
\int\!
A_{m+n-1,\,\mu+\nu\,}\,
\Big[\alpha_{m,\nu},\,\beta_{n,\mu}\Big]
%\bigg\rangle
{\rm\, d}^N{\bf q}
\\
&\hspace{0.63cm}+
\int\hspace{-3mm}\int\hspace{-3mm}\int\!
f\,{\bf p}^{m+n}\alpha_{m,\nu}({\bf q})\,\beta_{n,\mu}({\bf q})
\left(g_d\,C^d_{bc}\,\frac{\pa g_{a_1}\dots g_{a_\nu}}{\pa g_b}\,\frac{\pa g_{s_1}\dots g_{s_\mu}}{\pa
g_c}\right)
{\rm d}^N{\bf q}\wedge{\rm d}^N{\bf p}\wedge{\rm d}^K{\bg}
\\
=&
\int\!
A_{m+n-1,\,\mu+\nu\,}\,
\Big[\alpha_{m,\nu},\,\beta_{n,\mu}\Big]
%\bigg\rangle
{\rm\, d}^3{\bf q}
\\
&\hspace{0.63cm}+
\int\hspace{-3mm}\int\hspace{-3mm}\int\!
f\,{\bf p}^{m+n}\left(\alpha_{m,\nu}\right)^{a_1,\dots,a_{\nu-1},b}\,
\left(\beta_{n,\mu}\right)^{s_1,\dots,s_{\mu-1},c}
\\
&\hspace{4.8cm}
\Big(hk\,g_d\,\,C^d_{bc}\,\,g_{a_1}\dots g_{a_{\nu-1}}\,g_{s_1}\dots g_{s_{\mu-1}}\Big)
{\rm d}^N{\bf q}\wedge{\rm d}^N{\bf p}\wedge{\rm d}^K{\bg}
\\
=&
\int\!
A_{m+n-1,\,\mu+\nu\,}\,
\Big[\alpha_{m,\nu},\,\beta_{n,\mu}\Big]
%\bigg\rangle
{\rm\, d}^3{\bf q}
\\
&\hspace{0.63cm}+h\,k
\int\hspace{-3mm}\int\hspace{-3mm}\int\!
f\,{\bf p}^{m+n}
\Big(g_{a_1}\dots g_{a_{\mu+\nu-1}}\Big)
\\&
\hspace{3.87cm}
\,\left(\alpha_{m,h}\right)^{a_1,\dots,a_{\nu}}\,C^{^\text{\scriptsize\,$a_{\mu+\nu-1}$}}_{a_\nu\,\,a_{\mu+\nu}}\,
\left(\beta_{n,\mu}\right)^{a_{\nu+1},\dots,a_{\mu+\nu}}\,
{\rm d}^N{\bf q}\wedge{\rm d}^N{\bf p}\wedge{\rm d}^K{\bg}
\\
=&\,
\bigg\langle
A_{m+n-1,\,\mu+\nu\,},
\Big[\alpha_{m,\nu},\,\beta_{n,\mu}\Big]
\bigg\rangle
+\nu\,\mu\,\Big\langle
A_{\,m+n,\,\mu+\nu-1}\,,\,
C\contract \,\alpha_{m,\nu}\otimes\beta_{n,\mu}
\Big\rangle
\\
=&\!\!:\!
%\fbox{$\displaystyle
\bigg\langle
A_{m+n-1,\mu+\nu\,},
\Big[\alpha_{m,\nu},\,\beta_{n,\mu}\Big]
\bigg\rangle
+
\Big\langle
A_{\,m+n,\,\mu+\nu-1}\,,\,
\Big[\alpha_{m,\nu},\,\beta_{n,\mu}\Big]_{\!C}
%C\contract \,\alpha_{m,h}\otimes\beta_{n,k}
\Big\rangle
%$}
\end{align*}
which finally coincides with equation (\ref{GHK-Schouten}).

\bigskip

%%%%%%%%%%%%%%%%%%%%%
\bibliographystyle{unsrt}

%%%%%%%%%%%%%%%%%%%%%

\end{document}